\documentclass[journal=nalefd,manuscript=letter,layout=twocolumn]{achemso}
\setkeys{acs}{etalmode=truncate,maxauthors=10} 
\usepackage{graphicx}
\usepackage{dcolumn}
\usepackage{bm}
\usepackage{gensymb}
\usepackage{siunitx}
\usepackage{amsmath}
\usepackage{amssymb}

\newcommand{\dg}[1]{\ensuremath{#1^\circ}}
\newcommand{\wavenumber}[1]{\ensuremath{#1~\text{cm}^{-1}}}

\newcommand{\mumetr}[1]{\ensuremath{\SI{#1}{\micro\meter}}}
\newcommand{\nmetr}[1]{\ensuremath{#1~\text{nm}}}
\newcommand{\twotwoMatrix}[4]{\ensuremath{\begin{pmatrix}
#1 & #2 \\ 
#3 & #4 \end{pmatrix}}}

\author{Nikolai Christian Passler}
 \affiliation{Fritz-Haber-Institut der Max-Planck-Gesellschaft, Faradayweg 4-6,14195 Berlin, Germany}
 \email{passler@fhi-berlin.mpg.de}
\author{Christopher R. Gubbin}
 \affiliation{School of Physics and Astronomy, University of Southampton, Southampton SO17 1BJ, United Kingdom}
\author{Thomas Graeme Folland}
 \affiliation{Vanderbilt Institute of Nanoscale Science and Engineering, 2301 Vanderbilt Place, PMB 350106, Nashville, TN 37235-0106, USA}
\author{Ilya Razdolski}
 \affiliation{Fritz-Haber-Institut der Max-Planck-Gesellschaft, Faradayweg 4-6,14195 Berlin, Germany}
\author{D. Scott Katzer}
 \affiliation{US Naval Research Laboratory, 4555 Overlook Avenue SW, Washington DC 20375, USA}
\author{David F. Storm}
 \affiliation{US Naval Research Laboratory, 4555 Overlook Avenue SW, Washington DC 20375, USA}
\author{Martin Wolf}
 \affiliation{Fritz-Haber-Institut der Max-Planck-Gesellschaft, Faradayweg 4-6,14195 Berlin, Germany}
\author{Simone De Liberato}
 \affiliation{School of Physics and Astronomy, University of Southampton, Southampton SO17 1BJ, United Kingdom}
\author{Joshua D. Caldwell}
 \affiliation{Vanderbilt Institute of Nanoscale Science and Engineering, 2301 Vanderbilt Place, PMB 350106, Nashville, TN 37235-0106, USA} 
 \alsoaffiliation{US Naval Research Laboratory, 4555 Overlook Avenue SW, Washington DC 20375, USA}
\author{Alexander Paarmann}
 \affiliation{Fritz-Haber-Institut der Max-Planck-Gesellschaft, Faradayweg 4-6,14195 Berlin, Germany}
 \email{alexander.paarmann@fhi-berlin.mpg.de}

\title{Strong Coupling of Epsilon-Near-Zero Phonon Polaritons in Polar Dielectric Heterostructures}

\keywords{surface phonon polariton, epsilon near zero, infrared, nanophotonics, strong coupling, hybridization}

\begin{document}



%
%
%


\begin{abstract}
We report the first observation of epsilon near zero (ENZ) phonon polaritons in an ultrathin AlN film fully hybridized with surface phonon polaritons (SPhP) supported by the adjacent SiC substrate. Employing a strong coupling model for the analysis of the dispersion and electric field distribution in these hybridized modes, we show that they share the most prominent features of the two precursor modes. The novel ENZ-SPhP coupled polaritons with a highly propagative character and deeply sub-wavelength light confinement can be utilized as building blocks for future infrared and terahertz (THz) nanophotonic integration and communication devices.
\end{abstract}


Integrated THz photonics relies on the development of artificially designed nano-scale metamaterials, where subwavelength structures in periodic patterns enable precise tuning of the material's optical response\cite{Joannopoulos2008,Burgos2010,DeglInnocenti2018}. Truly extraordinary light propagation characteristics can be achieved in metamaterial-based epsilon near zero (ENZ) media\cite{Li2015,Liberal2017}, i.e. where the dielectric permittivity is vanishingly small. In particular, remarkable properties of the ENZ photonic modes include tunneling through narrow distorted channels\cite{Silveirinha2007,Edwards2009}, enhanced nonlinear-optical conversion efficiency via enforced phase-matching\cite{Argyropoulos2012,Suchowski2013,Mattiucci2014}, high emission directionality\cite{Enoch2002,Ziolkowski2004,Kim2016} and enable polaritonic waveguiding modes with broken time inversion symmetry and reduced scattering rate\cite{Liu2016,Engheta2007}.

An important challenge of future nanophotonics consists in enabling effective nanoscale communication and long-range information transfer. Both those objectives can be greatly facilitated by exploiting the unique properties of ENZ physics. In this regard, it is of key importance to find an appropriate class of systems where high-Q ENZ polaritons can be efficiently excited and coupled to other photonic excitations, while maintaining highly propagative character with reasonable group velocities. Commonly, bulk ENZ photonic modes excited in carefully designed metamaterials based on plasmonic nanostructures are considered, which are, however, characterized by high losses\cite{Drachev2008,Khurgin2012}. A complimentary approach aims at utilizing the naturally occuring zero-crossing of the dielectric function in the spectral vicinity of intrinsic material vibrations, such as the transverse optical (TO) and longitudinal optical (LO) phonons in polar dielectric crystals\cite{Campione2015,Nordin2017,Vassant2012}. Yet, employing freestanding films of polar dielectrics\cite{Nordin2017}, however, bears little practical importance, and the low dispersion results in a non-propagative character of the ENZ polaritonic modes\cite{Campione2015}.


\begin{figure}
\includegraphics[width=.9\linewidth]{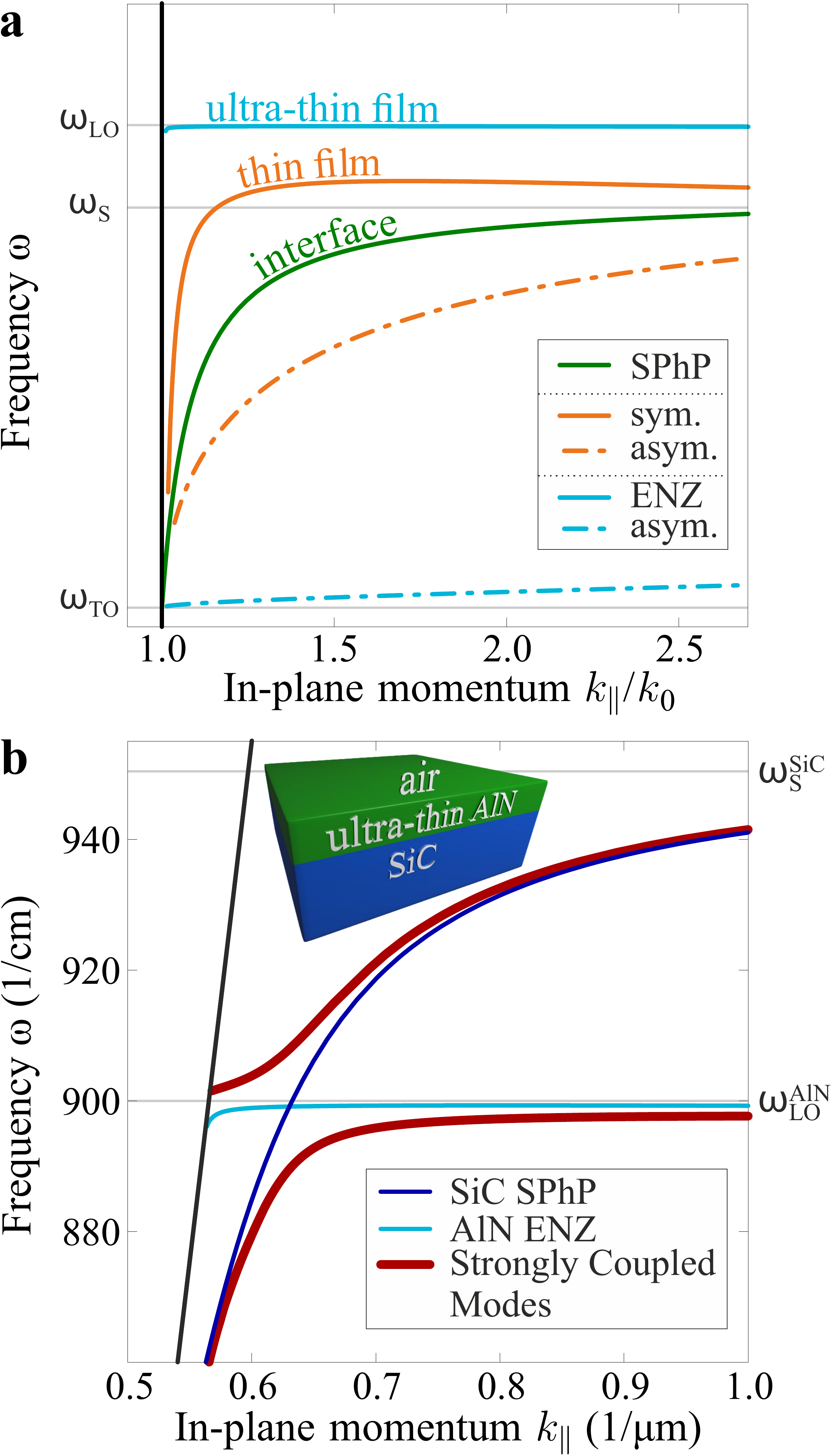}
  \caption{\textbf{Interaction of a surface phonon polariton with an epsilon-near-zero mode.} \textbf{a} A SPhP propagating at a single interface of air and a polar crystal has a dispersion relation (green curve) ranging from the transversal optical phonon frequency $\omega_{TO}$ up to a cut-off frequency $\omega_{S}$, shown in green. For film thicknesses $d<\lambda / 2$, where $\lambda$ is the free-space wavelength, a symmetric and an anti-symmetric branch appear (orange), splitting further apart with decreasing $d$. In the ultrathin limit of $d<\lambda / 100$, the upper branch is pushed close to the longitudinal optical phonon frequency $\omega_{LO}$ (light blue), where the permittivity exhibits a natural zero-crossing. Therefore, the symmetric branch is termed an epsilon-near-zero mode. \textbf{b} When an ultrathin AlN film is placed on bulk SiC, the AlN epsilon-near-zero mode (light blue) intersects the SiC SPhP dispersion relation (dark blue). These modes strongly interact, inducing an avoided crossing and forming two new dispersion branches of the coupled system, drawn in red.}
  \label{fig1}
\end{figure}


In this work, we suggest a novel concept for the hybrid ENZ excitations, which utilizes strong coupling between the ENZ and surface phonon polaritons (SPhPs) in the reststrahlen band of polar dielectrics (demarcated by the phonon frequencies $\omega_{TO}$ and $\omega_{LO}$ \cite{Caldwell2015,Feng2015}). We demonstrate the hybridization of an ENZ polariton and a propagating low-loss SPhP at an adjacent interface in the strong coupling regime, thus adding a new pair of coupled nanophotonic excitations to an evergrowing suite\cite{Caldwell2016,Basov2016,Low2017,Simpkins2015,Dunkelberger2016}. We analyze the coupling of the ENZ and SPhP modes in an AlN/SiC bilayer, where the ENZ polariton in AlN occurs within the reststrahlen band of SiC. The novel coupled ENZ-SPhP modes inherit their properties from both ENZ and SPhP components, thus enabling highly efficient phase-matched exitation. Offering broad functionality, these ENZ-SPhP coupled modes feature a unique combination of deeply subwavelength confinement, large enhancements of the local electromagnetic fields as well as an intrinsically low-loss, propagative character with non-zero group velocity. 

A SPhP mode supported at the interface of a polar dielectric in the Reststrahlen band is split into two branches upon reducing the film thickness $d$ (Fig. \ref{fig1}a), known as symmetric and antisymmetric modes\cite{Burke1986,Campione2015,Nordin2017}. Remarkably, in ultrathin films ($d/\lambda < 10^{-2}$, with $\lambda$ being the free-space wavelength) the upper (symmetric) mode is pushed towards the longitudinal optical (LO) phonon frequency\cite{Campione2015}, where the real part of the dielectric permittivity approaches zero ($\varepsilon^{\prime}(\omega=\omega_{\rm LO})=0$). While this ultrathin film ENZ polariton loses its dispersive character, its ultra-long wavelength leads to a strongly subwavelength mode confinement, enabling a gigantic enhancement of the electric field with minimal phase change over several times the free-space wavelength. The polariton dispersion curves in the bilayer with an ultrathin AlN film on top of a SiC substrate are exemplified in Fig. \ref{fig1}b. Individually, the AlN film exhibits a non-dispersive ENZ polariton mode (light blue) and the SiC a SPhP (dark blue), see Fig. S1a of the Supporting Information. The combined bilayer structure (inset Fig. \ref{fig1}b), however, reveals a strong interaction between the two modes, leading to two new dispersive branches featuring an avoided crossing (red). 

\begin{figure}
\includegraphics[width=.8\linewidth]{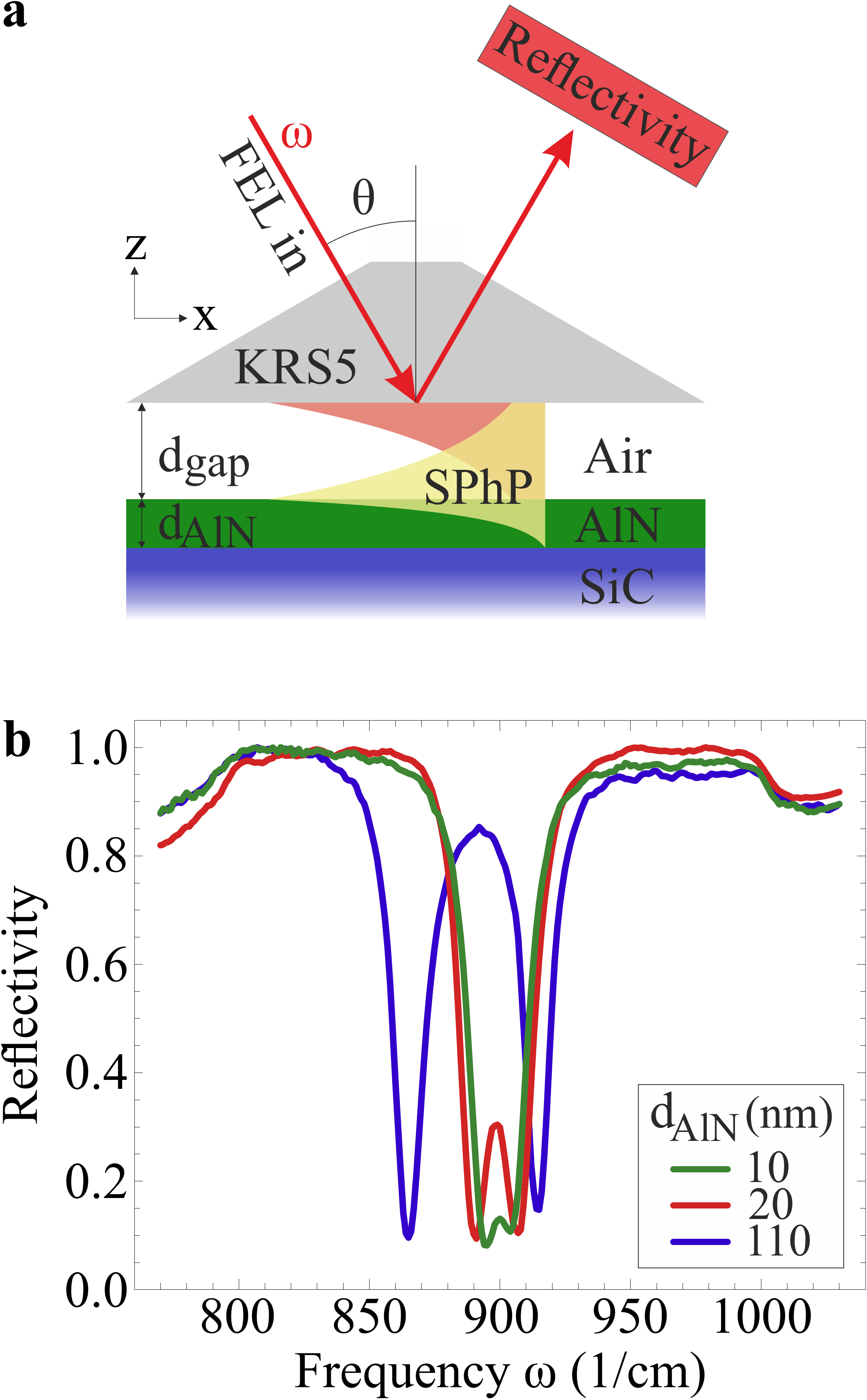}
  \caption{\textbf{The Rabi frequency dependence on the AlN film thickness.} \textbf{a} Prism coupling setup implementing the Otto configuration, where a highly-refractive KRS5 prism ($n_{KRS5}\approx 2.4$) enables phase-matched excitation of phonon polaritons across a variable air gap. By tuning the incoming frequency $\omega$ and the incidence angle $\theta$, polariton dispersion curves can be mapped out. \textbf{b} Reflectivity spectra for three different AlN film thicknesses $d_{AlN}=10,20,\nmetr{110}$ ($\theta = 29,29,\dg{30}$ and $d_{gap}=4.8,4.5,\mumetr{3.7}$, respectively). The resonance dips represent the two strongly interacting polariton branches in the AlN/SiC heterostructure.}
 \label{fig2a}
\end{figure}

\begin{figure*}[t]
\includegraphics[width=.95\linewidth]{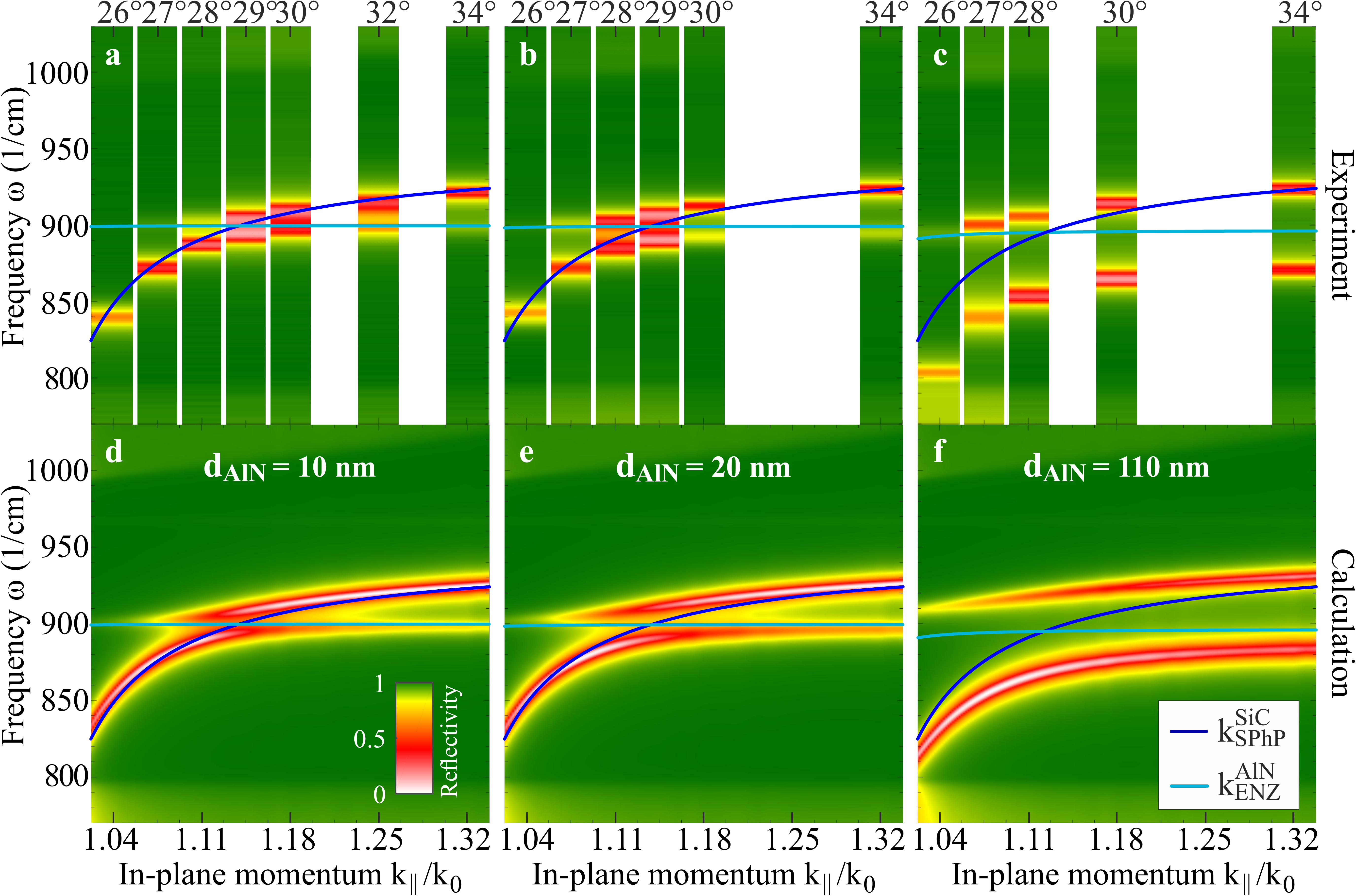}
  \caption{\textbf{Mapping out the strongly coupled polariton dispersion.} Experimental and calculated reflectivity maps for all three $d_{AlN}$, revealing the anticipated avoided crossing in the dispersion of the interacting polariton modes. The transfer matrix calculations (d-f) perfectly reproduce the experimental data (a-c). Clearly, the splitting of the modes is large for the thickest film, and decreases for smaller thicknesses. The dark and light blue lines indicate the dispersion of the uncoupled SPhP and ENZ modes, respectively. The former is calculated in the limit of vanishing AlN thickness, while the latter is the dispersion of the ultrathin film polariton in a freestanding AlN film. We note that for the here shown film thicknesses, the dependence of the dispersion of the ultrathin film mode on $d_{AlN}$ (compare to Fig. \ref{fig1}a) is negligible.}
 \label{fig2b}
\end{figure*}

In the experiments, we realize phase-matched excitation of the coupled modes by employing the Otto prism geometry\cite{Otto1968,Passler2017}, with a schematic provided in Fig. \ref{fig2a}a. For total internal reflection inside the prism, the evanescent wave at the prism backside enables phase-matched excitation of the polariton modes in the sample. Spectroscopic reflectivity measurements with varied incidence angle allow for the mapping out of the polariton dispersion relation (for details on the experimental methods see Supporting Information Section 1). In our AlN/SiC structure, two polariton branches are present, and hence two resonance dips can be observed (Fig. \ref{fig2a}b), showing that the frequency splitting of the branches increases for thicker films. This trend is also evident in the experimental reflectivity maps and the transfer matrix calculations in Fig. \ref{fig2b}. Furthermore, our data reveal the anticipated avoided crossing in the dispersion of the interacting polariton modes, corroborating the strong coupling mechanism.

\begin{figure*}
\includegraphics[width=\linewidth]{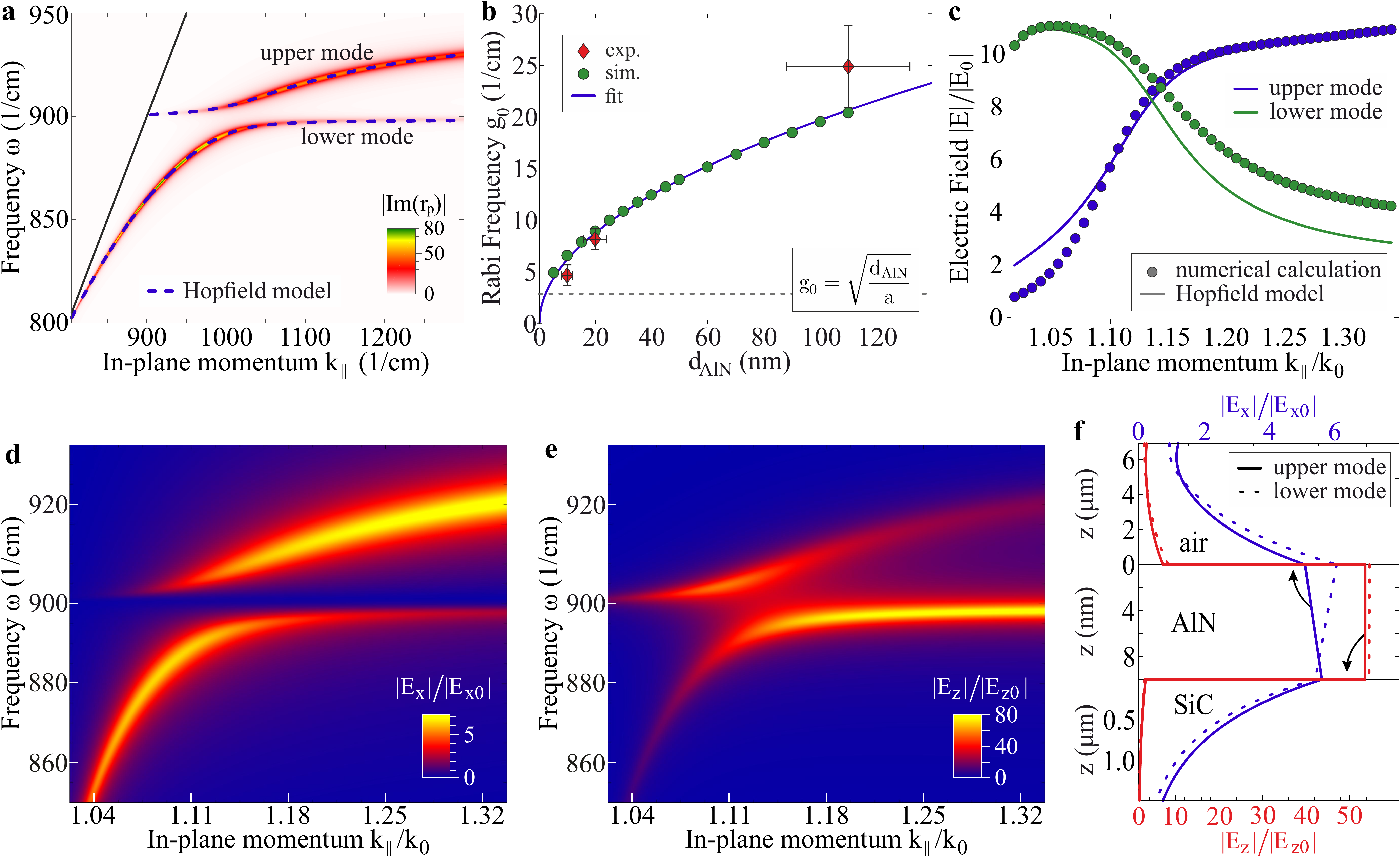}
  \caption{\textbf{Coupling between the ENZ and SPhP modes.} \textbf{a} Dispersion curves obtained by the analytical strong coupling model (dashed line) for an AlN film thickness of $d_{AlN}=\nmetr{10}$ on SiC on top of the numerically calculated dispersion curves obtained by computing the absolute imaginary part of the p-polarized reflection coefficient of the three layer system air/AlN/SiC. The analytical model features excellent agreement with the numerical calculations. \textbf{b} Rabi frequency $g_0$ as a function of $d_{AlN}$ extracted from the analytical model (green circles) and experiments (red diamonds). As can be seen from the simple model fit (solid line, $a=0.26\cdot 10^{-13}\text{m}^{3}$), the Rabi frequency follows a square root function characteristic for strong coupling. The dotted gray line indicates the threshold coupling strength for strong coupling according to the criterion discussed in the text. \textbf{c} Electric field strength obtained by the analytical model (solid line) and numerical calculations (circles) for the two strongly coupled modes, calculated at a probe point in air at the sample surface. At the avoided crossing, the two modes are completely hybridized, sharing equal field strength. At smaller in-plane momentum than at the avoided crossing, the lower branch has a larger field strength than the upper branch, and vice versa at larger momenta. The mode with larger field strength exhibits SPhP character, since at the specific probe point, the SPhP field dominates, while the ENZ polariton is localized inside the AlN film. \textbf{d,e} Normalized in-plane ($E_x$) and out-of-plane ($E_z$) field components along the entire dispersion. The SPhP has large in-plane and the ENZ polariton large out-of-plane fields, while the respective other component is small. This allows to track the mode nature (SPhP or ENZ polariton) of both dispersion branches, which is exchanged at the avoided crossing, along the entire momentum range. \textbf{f} Normalized $E_x$ and $E_z$ fields of the upper and the lower mode at resonance ($k_{\parallel}/k_0 = 1.13$) across the air/AlN/SiC structure. The layer thicknesses are not to scale with respect to each other.}
  \label{fig3}
\end{figure*}

Analytically, the strong coupling can be described by a system of two coupled oscillators, modeling the ENZ and the SPhP modes respectively. This allows us to calculate, in an analytically transparent way, the dispersion and the in-plane $E$-field distributions of the normal modes in the hybrid structure. The eigenfrequencies $\omega_q^{\pm}$ of the coupled system are then given by 

\begin{equation}
\omega_q^{\pm} = \frac{\omega_q^{e}+\omega_q^{s} \pm \sqrt{\left( \omega_q^{e} - \omega_q^{s} \right)^2 + 4 g_0^2}}{2},
\label{eq:freq}
\end{equation}

where $\omega_q^{\text{e, s}}$ are the bare frequencies of the ENZ and SPhP mode, respectively, and $g_0$ is the Rabi frequency, quantifying their mutual interaction. Although the eigenfrequencies in Eq. \ref{eq:freq} can be derived using a classical coupled mode approach, as described in the Methods section, we decided to use instead the Hopfield model usually employed in solid-state cavity quantum electrodynamics,\cite{Hopfield1958,Savona1994,Gubbin2016,Gubbin2016a} which in our opinion allows to gain a better insight of the hybrid nature of the resulting eigenmodes. We employ Eq.~\ref{eq:freq} to calculate the dispersion analytically (Fig. \ref{fig3}a), finding excellent agreement of our strong coupling model with the numerical calculations and demonstrating that the description of the energetic hybridization of the modes in terms of strong coupling is correct.


Having established the strongly coupled oscillators model for the ultrathin AlN films on the SiC substrate, we now turn to its examination as a function of the AlN film thickness $d_{AlN}$. Two effects can be identified, which play an important role in the evolution of the strong coupling as $d_{AlN}$ increases. First, since the SPhP is localized at the AlN/SiC interface and the ENZ mode in the entire AlN film, the spatial overlap of the two modes decreases, thus reducing the effective coupling strength. We can evaluate this through the distributions of the electric field calculated using the $4 \times 4$ transfer matrix formalism. These field distributions, exemplified for $d_{AlN}=\nmetr{110}$ in Fig. S2c (see Supporting Information Section 4), indicate diminishing effective coupling strengths at $d_{AlN} \approx \nmetr{50}$ and beyond, while at smaller AlN thicknesses, this effect remains rather marginal, see Fig. \ref{fig3}f. Second, since phonon polaritons are the collective excitations of atomic vibrations across individual bonds, increasing $d_{AlN}$ in our model effectively results in a correspondingly larger oscillator strength $f_{ENZ}$. At these small thicknesses $d_{AlN} \ll \lambda$, the optical absorption scales almost linearly with thickness, leading to a concomitant linear increase of the oscillator strength. Within the strong coupling formalism, the Rabi frequency $g_0$ (determining half the distance between the two spectral peaks corresponding to the coupled modes) scales as $\sqrt{f_{ENZ}}$\cite{Lidzey1998,Cade2009,Baieva2012} and thus as $\sqrt{d_{AlN}}$. The Rabi frequency obtained from the fits of the calculated data using Eq.~\ref{eq:freq} follows this expected square root dependence with great accuracy, see Fig.~\ref{fig3}b. This behavior is further reproduced by the experimental coupling strengths also shown in Fig. \ref{fig3}b, determined by half the frequency splitting in the respective reflectivity spectra (Fig. \ref{fig2a}b). Even though the Rabi frequency follows the $\sqrt{d_{AlN}}$ dependence up to $d_{AlN}\approx \nmetr{500}$, the effective coupling strength is maximal for $d_{AlN} < \nmetr{50}$ due to the aforementioned loss of the spatial overlap of the two modes.

It is thus seen that the AlN/SiC bilayer can be tuned from the weak into the strong coupling regime by modifying the AlN layer thickness. In order to determine the lower limit of the $d_{AlN}$ range of strong coupling, we employ the criterion that, quite intuitively, the energy exchange rate between the two strongly coupled oscillators should exceed the loss rate, resulting in the appearance of two distinct frequencies in the spectrum\cite{Auffeves2010,Rodriguez2016}. A reliable and sufficient indication of the strong coupling is thus the avoided resonance crossing behavior (Fig. \ref{fig2b}) occurring if $2g_0/\gamma^* > 1$ is fulfilled, where $\gamma^*=(\gamma^e+\gamma^s)/2$ is the average of the loss rates of the two oscillators.


For our system, we obtain an estimate for the average loss rate of $\gamma^*=5.8$.\footnote{While $\gamma^s$ can be readily calculated using the transfer matrix method, determination of $\gamma^e$ is not straightforward. As an estimation, we assume the proportionality of the loss rate $\gamma$ to the imaginary part of the permittivity $\text{Im}(\varepsilon)$ of the corresponding material where the mode is largely localized (AlN and SiC for the ENZ and SPhP modes, respectively). As such, at $\omega \approx \wavenumber{900}$ we get $\gamma^s \approx \wavenumber{8.7}$, $\gamma^e \approx \wavenumber{3.0}$, and thus $\gamma* \approx \wavenumber{5.8}$.} The horizontal dotted line in Fig. \ref{fig3}b illustrates the threshold coupling strength (\wavenumber{2.9}) corresponding to the onset of the strong coupling regime according to the aforementioned criterion. This value can be reached in AlN films with thicknesses of \nmetr{2.2}. It is therefore seen that in the ultrathin AlN films on a SiC substrate discussed in this work, the ENZ and SPhP polaritons can indeed be strongly coupled for AlN film thicknesses $\nmetr{2.2} \lesssim d_{AlN} \lesssim \nmetr{50}$. 


Furthermore, our simple analytical model also correctly describes the electric field profiles. This is illustrated in Fig. \ref{fig3}c, where we show the coinciding analytical and numerically calculated field intensities of both modes in front of the sample. We note that while a bare SiC substrate allows the SPhP component of the coupled modes to be quantified, the ENZ polariton component depends decisively on the substrate material and hence cannot be straightforwadly quantified. Therefore, we assumed the ENZ polariton to be fully confined in the AlN film, i.e., the field at the probe point is solely determined by the SPhP component of the coupled modes. This assumption is the reason for the discrepancy between the analytical and calculated field intensities at in-plane momenta where the respective mode features ENZ character ($k_{\parallel}/k_0 < 1.1$ and $k_{\parallel}/k_0 > 1.2$ in Fig. \ref{fig3}c). However, despite its simplicity, our model reproduces the numerical field amplitudes extremely well, proving that the coupled modes can be described as a linear superposition of the ENZ and SPhP modes, weighted by the Hopfield coefficients. In consequence of this linear relationship, the strongly coupled modes at the avoided crossing share equal weights of SPhP and ENZ character, while the respective partitions change along the dispersion: the lower polariton starts as pure SPhP at small $k$ and switches to ENZ beyond the avoided crossing, while the upper polariton shows the opposite behavior. 

This switching of the mode nature can be illustrated by means of the in-plane ($E_x$) and out-of-plane ($E_z$) electric field components inside the AlN film, shown in Fig. \ref{fig3}d and e, respectively. Note that the SPhP is characterized by a large in-plane field, whereas the ENZ polariton features pronounced out-of-plane field enhancement. The lower branch has strong in-plane fields at lower momentum (Fig. \ref{fig3}d), illustrating that the mode is predominantly SPhP in nature. In contrast, the upper branch exhibits strong out-of-plane character at low $k_{\parallel}$ (Fig. \ref{fig3}e). Across the strong coupling region, the modes interchange these characteristics, with the upper branch exhibiting strong in-plane and the lower out-of-plane fields. At the avoided crossing, the fields of both modes are apparent and of equal weight, and even the spatial $E_x$ and $E_z$ field distributions of the modes across the multilayer structure show high agreement (Fig. \ref{fig3}f). We have thus demonstrated the strong coupling and full hybridization of an ultrathin film ENZ phonon polariton with a SPhP in a polar dielectric heterostructure exemplified for AlN/SiC. However, we emphasize that strong coupling will emerge for a large number of hybrid systems that feature overlapping reststrahlen bands of the two constituents\cite{Caldwell2015,Feng2015}. Furthermore, strong coupling can also be observed in the inverse structure of an ultrathin SiC film on AlN, ocurring at the TO frequency of SiC (see Supporting Information Section 3). 

Another important consideration which can be inferred from the distributions of the electric field is pertinent to the upper limit of the strong coupling. As mentioned above, for large thicknesses the spatial mismatch of the field distributions for the ENZ and SPhP modes (localized in the entire AlN film and at the AlN-SiC interface, respectively) becomes more and more important. For instance, in Supporting Information Section 4 the field distributions for the \nmetr{110} thick AlN film is exemplified, clearly showing the lack of full hybridization of the ENZ and SPhP modes. Our calculations reveal that this effect becomes significant for thicknesses $d_{AlN} \gtrsim \nmetr{50}$, thus reducing the effective coupling constant $g_0$. As such, we conclude that despite the fact that the mode splitting follows the $\sqrt{d_{AlN}}$ dependence up to $d_{AlN} \approx \nmetr{500}$, ultrathin films with deeply subwavelength thicknesses of $d\sim\lambda/1000$ ($\nmetr{11}$ for AlN) demonstrate the highest degree of strong coupling. In other words, within the strong coupling regime, thinner films provide higher quality ENZ wave characteristics, yet featuring full hybridization with the low-loss, highly confined SPhP, see Fig.~\ref{fig3}a-c. Within this regime, e.g. for $d_{AlN}=\nmetr{20}$, a propagation length of $L \approx \mumetr{900}$ with a group velocity of $v_g \approx 0.1 \text{c}$ of the coupled modes is achieved. Thereby, the ENZ-SPhP modes offer a new promising approach for THz photonics on the nanoscale using traditional materials like III-V and II-VI semiconductors\cite{Feng2015}. For instance, PbSe/PbS core-shell nanostructures\cite{Lifshitz2006} with wide tunability of the optical properties in the near-infrared additionally feature strongly coupled ENZ-SPhP modes in the THz range, allowing for a unique multispectral photonic integration. 

\begin{figure*}
\includegraphics[width=\linewidth]{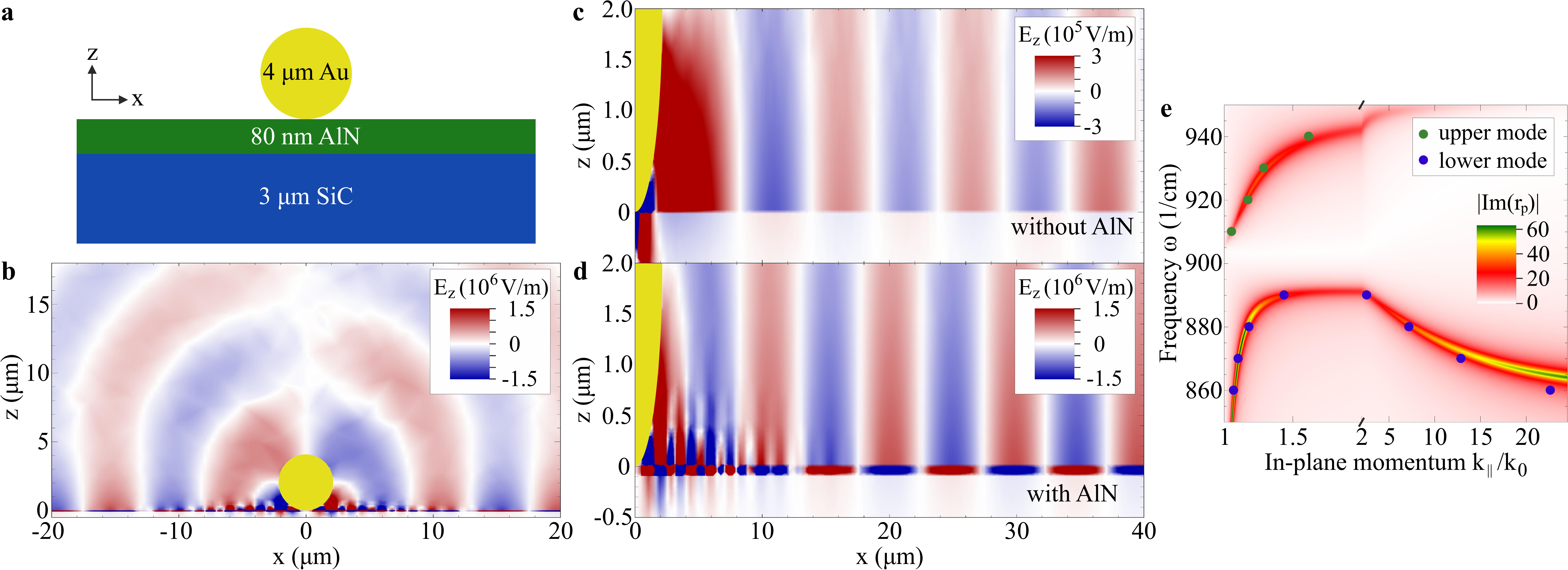}
  \caption{\textbf{Numerical simulations of a high-k polariton with negative group velocity.} \textbf{a} shows the simulation geometry, where plane waves are normally incident on a gold particle on the surface. Due to the large size of the particle, in \textbf{b} both surface waves and scattered free space waves are observed. \textbf{c,d} Surface electromagnetic field for a bare SiC substrate (c) and with an 80 nm AlN film on top (d). The field distributions in b-d were all calculated at an excitation frequency of \wavenumber{880}. \textbf{e} Comparison between the dispersion relation calculated by a transfer matrix approach (background color map) and the values extracted from the simulations (circles). At large in-plane momentum ($k_{\parallel}/k_0>2$), the negative slope of the dispersion of the lower mode reveals its negative group velocity.}
  \label{fig4}
\end{figure*}

To demonstrate that the general character of strong coupling is not restricted to the prism-based experiments, we complement our results with electromagnetic simulations of optical near-field scattering at nanoparticles. Motivated by previous results on scattering type scanning near field optical microscopy (s-SNOM) of Au nanostructures on SiC and boron nitride\cite{Huber2008, Dai2017}, we consider here a Au cylinder on top of the AlN film on a SiC substrate, see Fig. \ref{fig4}a. In the simulations, we monitor the spatial distribution of the electric field induced by normally incident plane waves at a series of frequencies. Whilst an \nmetr{80} AlN film was chosen in this simulation for a clear identification of the coupled polaritonic modes, similar results can be obtained for the AlN film thicknesses discussed earlier in this work.

Results for an \nmetr{80} AlN film on top of a SiC substrate are presented in Fig. \ref{fig4}b, showing the full electromagnetic field scattered from the Au cylinder at \wavenumber{880}. To focus on the polaritons launched along the surface, in Fig. \ref{fig4}c and d we consider the spatial distribution of the electric field along the surface in the two cases, namely, with and without the AlN layer. Without the AlN film (Fig. \ref{fig4}c) the results of the simulations are indicative of a propagating SPhP wave launched across the surface with minimal confinement. In the presence of the AlN film, however, the simulations reveal two distinct modes with unequal wavelengths, see Fig. \ref{fig4}d. The longer wavelength mode corresponds to that observed in our experiments, which is slightly compressed when compared with the wave propagating on the SiC surface. The short wavelength mode, however, was not observed in our measurements due to its large in-plane momentum. We attribute this mode to the breakdown of the thin film approximation for ENZ behaviour which has been theoretically predicted at extremely large k \cite{Campione2015}. One of the most striking features of this large-k mode is its negative group velocity, clearly visible from the dispersion in Fig. \ref{fig4}e and confirmed by the time-dependent E-field distributions (Movie S1).

To better understand both the positive and negative group velocity modes observed in these simulations, we quantify the polariton wavelength by Fourier transform (FT) of the simulated electromagnetic field. Here, we take the FT of the normal projection of the complex $E_z$ field on the top surface of the AlN layer at a series of different excitation frequencies. The dispersion determined from the $E_z$ Fourier spectra are plotted and compared against transfer matrix simulations of the polariton dispersion in Fig. \ref{fig4}e. The excellent agreement between the numerically simulated frequencies and the calculated dispersion indicates that the key results of this paper regarding the spatial distribution of fields should be observable by the s-SNOM technique. Additionally, the latter should enable the observation of the high-k mode with negative dispersion, which is otherwise inaccessible in the prism-coupling experiments. We thus envisage rich perspectives of near-field microscopy in visualizing coupled phonon-polaritonic modes in hybrid or multilayer systems.

In conclusion, in this work we have demonstrated and characterized polaritonic modes in a strong coupling regime between an ENZ polariton and a bulk SiC SPhP in an ultrathin AlN/SiC structure. The full mode hybridization at the avoided crossing enables unique propagating ENZ polaritons. We have performed numerical simulations, revealing that the s-SNOM approach enables the observation of both the propagation length of the coupled ENZ-SPhP modes, and the properties of polaritons featuring negative group velocity. Our results illustrate the high suitability of near-field techniques like s-SNOM for the investigation of low-loss ENZ polaritons in polar dielectric heterostructures, in order to further establish their potential for nanophotonic applications. We envision the generalization of employing polar dielectric ENZ heterostructures to open up a new platform of deeply sub-wavelength integrated THz photonics based on strongly coupled ENZ-SPhPs.

\section{Methods}

\subsection{Experimental}
The substrate of our samples is hexagonal 6H-SiC for the \nmetr{110} AlN film, and 4H-SiC for the other two samples, all three with the extraordinary axis perpendicular to the sample ($c$-cut). The AlN layers were grown by RF-plasma assisted molecular beam epitaxy, and therefore also exhibit a $c$-cut, hexagonal crystal structure.

As an excitation source, we employ a mid-infrared free electron laser (FEL) with small bandwidth ($\sim 0.3\%$) and wide tunability of $3-\mumetr{50}$, covering the spectral ranges of the SiC and AlN reststrahlen bands (details on the FEL have been reported elsewhere\cite{Schollkopf2015}). While the frequency is scanned by tuning the FEL, different in-plane momenta can be accessed via the incidence angle $\theta$ by rotating the entire Otto geometry (see Supporting Information Section 1 for more details), thus allowing for mapping out the complete dispersion curves experimentally\cite{Passler2017}. In contrast to alternative approaches, the Otto geometry features experimental control over the excitation efficiency through tunability of the air gap width $d_{gap}$. At each incidence angle, spectra were taken at several $d_{gap}$. For the reconstruction of the dispersion curves (Fig. \ref{fig2b}), we selected the spectra at a gap size of critical coupling conditions $d_{crit}$\cite{Passler2017}, i.e. where the polariton is excited the most efficiently (Supporting Information Section 2). Direct read-out of the gap width $d_{gap}$ with a range of $d=1-\mumetr{50}$ is realized via whitelight interferometry, while the contrast of the interference spectrum grants parallel alignment of prism and sample.
 
\subsection{Theoretical}
\subsubsection{Transfer Matrix}
All calculations of the optical response and field distributions of Fig. \ref{fig1}-\ref{fig3} were performed using a generalized $4 \times 4$ transfer matrix formalism\cite{Passler2017a}. In short, the formalism allows for the calculation of reflection and transmission coefficients in any number of stratified media with arbitrary dielectric tensor, which allows to account for the anisotropy of our samples. 

\subsubsection{Three-Layer Dispersion}
The dispersion curves in Fig. \ref{fig1} were obtained by numerical evaluation of the three-layer polariton dispersion formula \cite{Raether1988,Burke1986,Campione2015}
\begin{align}
1+\frac{\varepsilon_1 k_{z3}}{\varepsilon_3 k_{z1}}=i \tan{(k_{z2} d)} \left(\frac{\varepsilon_2 k_{z3}}{\epsilon_3 k_{z2}} + \frac{\varepsilon_1 k_{z2}}{\varepsilon_2 k_{z1}}  \right),
\label{threelayerdisp}
\end{align}

where the subscripts $i=1,2,3$ correspond to the three stacked media, $\varepsilon$ is the dielectric function, $d$ the film thickness of material $2$, $k_{zi}=\sqrt{\frac{\omega^2}{c^2} \varepsilon_{i} - k_{\parallel}^2}$ the out-of-plane momentum, and $k_{\parallel}$ the in-plane momentum conserved in all layers. 

\subsubsection{Hopfield Model}
In the rotating wave approximation, the two-oscillator Hopfield Hamiltonian for our system takes the form\cite{Hopfield1958,Savona1994,Gubbin2016,Gubbin2016a}
\begin{equation}
\mathcal{H} = \sum_q \hbar \omega_q^{\text{e}} \hat{a}_q^\dagger \hat{a}_q^{} + \hbar \omega_q^{\text{s}} \hat{b}_q^\dagger \hat{b}_q^{} + \hbar g_0 \left( \hat{a}_q^\dagger \hat{b}_q^{} + \hat{a}_q^{} \hat{b}_q^\dagger  \right),
\label{eq:ham}
\end{equation}

where $\hat{a}_q^{\dagger}$ ($\hat{a}_q^{}$) and $\hat{b}_q^{\dagger}$ ($\hat{b}_q^{}$) are the bosonic creation (annihilation) operators for the ENZ and SPhP modes. The Rabi frequency $g_0$ of the strong coupling model in Eq. \ref{eq:ham} is introduced as a phenomenological coupling parameter, and is equivalent to the overlap of the substrate SPhP and ENZ polariton in a classical electromagnetic approach. The eigenfrequencies (Eq. \ref{eq:freq}) of the coupled system are found by diagonalization of the Hopfield-Bogoliubov matrix $H_q$\cite{Gubbin2016}
\begin{align}
H_q=\twotwoMatrix{\omega_q^e}{g_0}{g_0}{\omega_q^s}
\end{align}

for each in-plane wavevector $q$ individually, where $e$ and $s$ stands for the ENZ polariton and the substrate SPhP, respectively. The eigenvalues of these matrices yield the eigenfrequencies $\omega_q^{\pm}$ shown in Eq. \ref{eq:freq}, and the respective normalized eigenvectors $\left( X_q, Y_q \right)$ are built from the Hopfield coefficients $X_q$ and $Y_q$, describing the weighting factors of the ENZ polariton and the substrate SPhP which compose the two hybridized modes along the avoided crossing. The analytic electric field strength shown in Fig. \ref{fig3}c has been calculated by multiplying these Hopfield coefficients with the electric field of the substrate SPhP at the probe point in air at the sample surface. The bosonic annihilation operators $\hat{p}_q$ of the coupled modes are then given by
\begin{align}
\begin{split}
\hat{p}_q^+ &= X_q \hat{a}_q + Y_q \hat{b}_q \\
\hat{p}_q^- &= Y_q \hat{a}_q - X_q \hat{b}_q,
\end{split}
\end{align}

where the superscripts $+$ and $-$ denote the upper and the lower coupled polariton branch, respectively.

\subsubsection{CST Simulations}
Simulations for Fig. \ref{fig4} were performed in CST studio suite\cite{CST} using the frequency domain solver. To approximate the structure shown in Fig. \ref{fig4}a within a finite 3D model, a unit cell with a size of \mumetr{250} by \mumetr{0.6} was chosen, which minimized nearest neighbour interactions. The optical constants of the respective materials were taken from literature\cite{Engelbrecht1993,Moore2005}. Unit cell boundaries were used at the edges of the substrate, and a matched impedance layer was used to suppress substrate reflections. Fourier analysis was performed using a one dimensional field profile running along the top surface of the ENZ film down the center of the unit cell.

\begin{acknowledgement}

We thank Wieland Sch\"ollkopf and Sandy Gewinner for operating the FEL. D.S.K., D.F.S, and J.D.C. were supported by the Office of Naval Research through the U.S. Naval Research Laboratory and administered by the NRL Nanoscience Institute. The NRL team acknowledges the AlN characterization and processing contributions of Neeraj Nepal, Brian P. Downey and Neil P. Green. S. D. L. is a Royal Society Research Fellow and he acknowledges support from the Innovation Fund of the EPSRC Programme EP/M009122/1.

\end{acknowledgement}

\begin{suppinfo}

Experimental details (Section 1), critical coupling conditions of the strongly coupled modes (Section 2), strongly interacting modes in materials with overlapping reststrahlen bands (Section 3), deviation from strong coupling for larger film thicknesses (Section 4), and simulated time-dependent E-field distributions (Movie S1).

\end{suppinfo}

\bibliography{strongCoupling}

\end{document}


\title{Supporting Information: \\ Strong Coupling of Epsilon-Near-Zero Phonon Polaritons in Polar Dielectric Heterostructures}

\author{Nikolai Christian Passler}
 \email{passler@fhi-berlin.mpg.de}
 \affiliation{Fritz-Haber-Institut der Max-Planck-Gesellschaft, Faradayweg 4-6,14195 Berlin, Germany}

\author{Christopher R. Gubbin}
 \affiliation{School of Physics and Astronomy, University of Southampton, Southampton SO17 1BJ, United Kingdom}
\author{Thomas Graeme Folland}
 \affiliation{Vanderbilt Institute of Nanoscale Science and Engineering, 2301 Vanderbilt Place, PMB 350106, Nashville, TN 37235-0106}
\author{Ilya Razdolski}
 \affiliation{Fritz-Haber-Institut der Max-Planck-Gesellschaft, Faradayweg 4-6,14195 Berlin, Germany}
\author{D. Scott Katzer}
 \affiliation{US Naval Research Laboratory, 4555 Overlook Avenue SW, Washington DC 20375, USA}
\author{David F. Storm}
 \affiliation{US Naval Research Laboratory, 4555 Overlook Avenue SW, Washington DC 20375, USA}
\author{Martin Wolf}
 \affiliation{Fritz-Haber-Institut der Max-Planck-Gesellschaft, Faradayweg 4-6,14195 Berlin, Germany}
\author{Simone De Liberato}
 \affiliation{School of Physics and Astronomy, University of Southampton, Southampton SO17 1BJ, United Kingdom}
\author{Joshua D. Caldwell}
 \affiliation{Vanderbilt Institute of Nanoscale Science and Engineering, 2301 Vanderbilt Place, PMB 350106, Nashville, TN 37235-0106} 
 \affiliation{US Naval Research Laboratory, 4555 Overlook Avenue SW, Washington DC 20375, USA}
\author{Alexander Paarmann}
 \email{alexander.paarmann@fhi-berlin.mpg.de}
 \affiliation{Fritz-Haber-Institut der Max-Planck-Gesellschaft, Faradayweg 4-6,14195 Berlin, Germany}

\date{\today}
\maketitle
\beginsupplement

\section{Experimental}

The experimental setup is sketched in Fig. 2a of the main text. The Otto geometry is implemented using three motorized actuators (Newport TRA12PPD) to control the position of the triangular coupling prism (KRS5, 25mm high, 25mm wide, angles of \dg{30}, \dg{30}, and \dg{120}, with the \dg{120} edge cut-off, Korth Kristalle GmbH) relative to the sample. The motors push the prism against springs away from the sample and allow for continuous tuning of the air gap $d_{gap}$. The \dg{120} prism edge is cut-off parallel to the backside in order to perpendicularly couple in the collimated light from a stabilized broadband tungsten-halogen light source (Thorlabs SLS201L). The back-reflected spectrum (registered with Ocean Optics, NIRQuest512) carries interferometric information about the air gap, allowing to quantify $d_{gap}$ by means of the modulation period, and to achieve parallel alignment of prism and sample by optimizing the contrast of the spectral modulations. The absolute read-out of $d_{gap}$ is limited to maximally $\sim\mumetr{60}$ due to the resolution of the spectrometer, while gap sizes below \mumetr{1} could be measured in principle, but are typically not feasible because of macroscopic protrusions and non-flatness of the sample.

At external incidence angles $\theta_{ext}$ away from \dg{30}, the high refractive index of KRS5 ($n_{KRS5}\approx 2.4$) leads to strong refraction, resulting in an internal incidence angle $\theta$ inside the prism of $\theta=\dg{30}+\arcsin \left( \sin \left( \theta_{ext}-\dg{30} \right) / n_{KRS5} \right)$. In this configuration, angles from below the critical angle of total internal reflection ($\theta_{crit}\approx \dg{25}$) up to about \dg{52} are accessible, covering in-plane momenta across and far beyond the anti-crossing of the strongly coupled modes (\dg{52} corresponds to $k_{\parallel}/k_0=1.86$). Different in-plane momenta can be accessed via the incidence angle $\theta$ by rotating the entire Otto geometry, thus allowing for mapping out the complete dispersion curves experimentally.

As excitation source we employ a mid-infrared free electron laser (FEL) as tunable, narrowband, and p-polarized excitation source. Details on the FEL are given elsewhere\cite{Schollkopf2015}. In short, the electron gun is operated at a micropulse repetition rate of $1$ GHz with a macropulse duration of \musec{10} and a repetition rate of $10$ Hz. The electron energy is set to 31 MeV, allowing to tune the FEL output wavelength between $\sim \mumetr{7-18}$ ($\sim \wavenumber{1400-550}$, covering the spectral ranges of the SiC and AlN reststrahlen bands) using the motorized undulator gap, providing ps-pulses with typical full-width-at-half-maximum (FWHM) of $<\wavenumber{5}$. The macropulse and micropulse energies are $\sim30$ mJ and $\sim10~\text{\miu J}$, respectively.

The reflectivity signal is detected with a pyroelectric sensor (Eltec 420M7). A second sensor is employed to measure the power reflected off a thin KBr plate placed prior to the setup in the beam. This signal is used as reference for normalization of all reflectivity signals, and is measured on a single shot basis in order to minimize the impact of shot-to-shot fluctuations of the FEL. The spectral response function of the setup is determined by measuring the reflectivity at large gap widths $d_{gap}\approx \mumetr{40}$, where the prism features total internal reflection across the full spectral range. This is done for each incidence angle individually, and all spectra are normalized to their respective response function. 

\begin{figure*}[t]
\includegraphics[width=.84\linewidth]{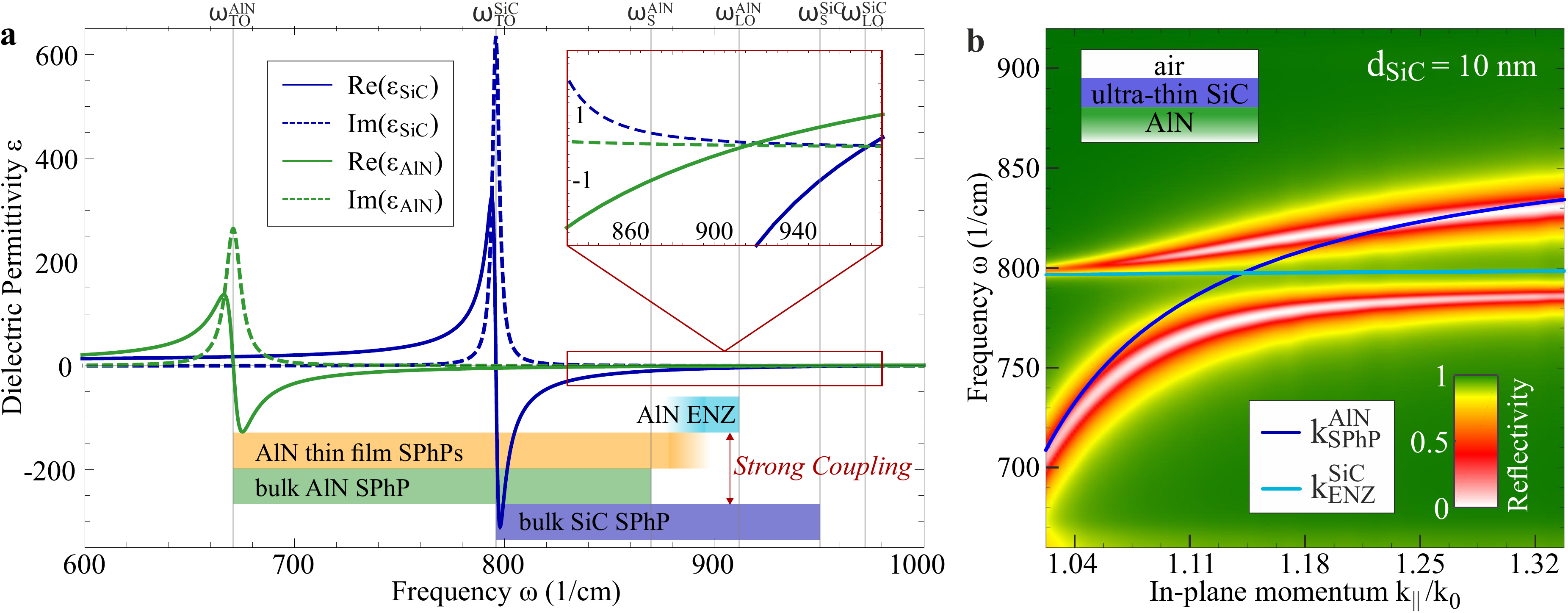}
  \caption{\textbf{Frequency ranges of phonon polaritons in AlN and SiC.} \textbf{a} The real (solid line) and imaginary (dashed line) part of the dielectric function $\varepsilon$ of SiC (blue) and AlN (green) are shown. At the respective optically active TO phonon frequency $\omega_{TO}$, they feature a strong resonance, while at the LO frequencies $\omega_{LO}$, the real part of $\varepsilon$ crosses zero, as illustrated in the inset. In a bulk material, a single SPhP is supported in the frequency range between $\omega_{TO}$ and the cut-off frequency $\omega_S$ (blue and green color bars). In a thin AlN film ($d_{AlN}<\lambda/2$), two thin film polaritons emerge with the asymmetric mode having a dispersion at frequencies above $\omega_S$ (orange color bar), see main text. In the ultra-thin limit of an AlN film ($d_{AlN}<\lambda/50$), the asymmetric mode disperses close to $\omega_{LO}^{AlN}$ where $Re(\varepsilon_{AlN})\approx 0$, hence being an ENZ mode. The strong coupling demonstrated in the main text arises between the SiC substrate SPhP and the AlN ENZ mode. \textbf{b} Reflectivity map of the inverse structure of an ultrathin SiC film on top of an AlN substrate at frequencies around the TO frequency of SiC $\omega_{TO}^{SiC}\approx \wavenumber{796}$. The theoretical dispersions of the asymmetric SiC ENZ mode (light blue line) and the substrate AlN SPhP (dark blue line) are plotted on top, exhibiting a crossing point analogous to the structure discussed in the main text. Interestingly, the dispersion in the combined structure also splits up into two branches, featuring an avoided crossing.}
  \label{sfig1}
\end{figure*}

\section{Critical Coupling Conditions of the Strongly Coupled Modes}

The Otto geometry grants experimental control over the excitation efficiency by means of the tunability of the air gap width $d_{gap}$. At a certain critical gap $d_{crit}$, the excitation efficiency is optimal \cite{Passler2017}, and we emphasize that $d_{crit}$ is significantly different for the two modes considered ($d_{crit}^{ENZ}<d_{crit}^{SPhP}$). Therefore, in a single spectrum with fixed gap width, the two modes cannot be excited equally efficiently, resulting in drastically different resonance dip depths. The data shown in Fig. 2b of the main text, however, were taken at in-plane momentum corresponding to the strong coupling conditions. Note that the two reflectivity dips in each spectrum are of equal depth, evidencing the hybrid nature of the two modes. The tunability and reproducibility of the gap in our setup is therefore a substantial tool for the measurement of polariton resonances, revealing information about their coupling efficiency.

\section{Strongly interacting Modes in Materials with Overlapping Reststrahlen Bands}

\begin{figure*}
\includegraphics[width=\linewidth]{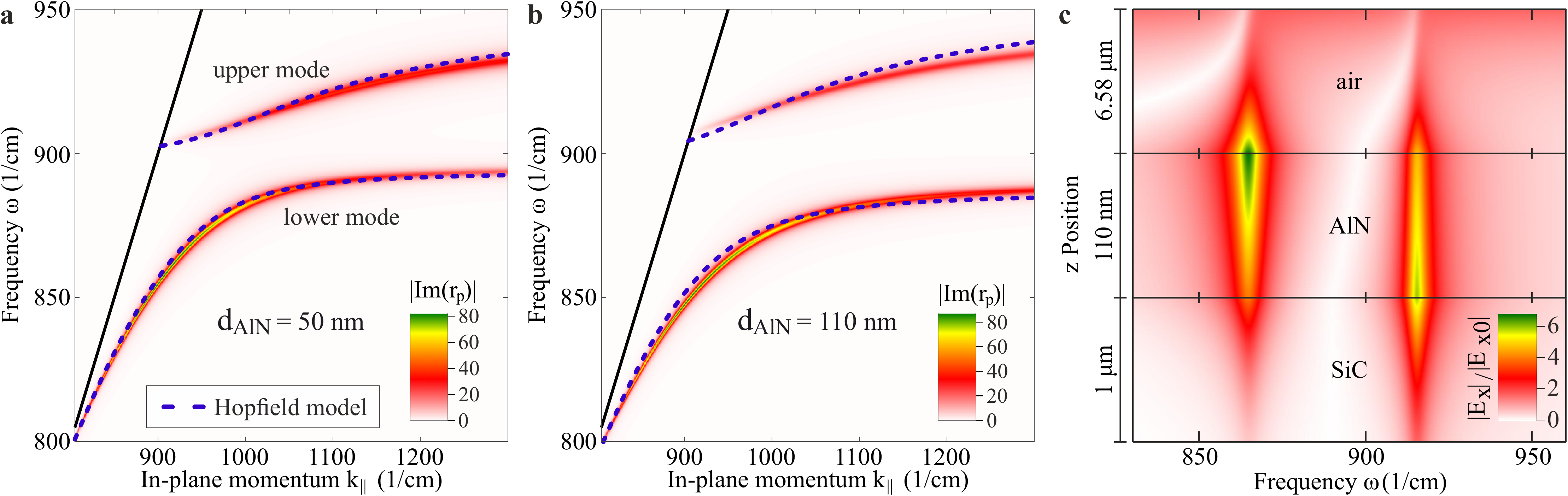}
  \caption{\textbf{Deviation from strong coupling for larger film thicknesses.} \textbf{a,b} Reflectivity maps of the AlN/SiC structure from the main text in the Otto geometry for $d_{AlN}=\nmetr{50,110}$, respectively, featuring the two dispersion branches of the interacting modes with avoided crossing. On top, the dispersion curves extracted from the strong coupling Hopfield model are plotted (blue dashed lines), exhibiting increasing deviation from the numerical calculations for larger film thicknesses. \textbf{c} In-plane ($x$) field enhancement $E_x$ as a function of frequency and $z$-position plotted at the avoided crossing and for an AlN film thickness of $d_{AlN}=\nmetr{110}$, where the layer thicknesses are not to scale with respect to each other. The prism lies adjacent to air at the top, but is not shown here. At the corresponding resonance frequency of the two modes, strong field enhancements can be observed, which peak at the air/AlN and the AlN/SiC interface, demonstrating that the two modes are not hybridized anymore.}
  \label{sfig2}
\end{figure*}

In Fig. \ref{sfig1}a, the dielectric functions $\varepsilon$ of AlN and SiC are shown. They feature a strong resonance at the respective TO frequency $\omega_{TO}$ due to its infrared activity, while at the non-IR-active LO frequency $\omega_{LO}$, the real part of $\varepsilon$ crosses zero. For the strong coupling to arise, it is necessary that the reststrahlen bands of the two materials overlap, i.e. in the case of an AlN film on SiC, that the $\omega_{LO}$ frequency of AlN is located inside the reststrahlen band of SiC. Only then, the ENZ mode supported by the ultra-thin AlN film lies at frequencies where the SiC substrate supports a SPhP, allowing the modes to strongly interact, as demonstrated in the main text.

A similar scenario is obtained for the inverse structure not discussed in the main text, i.e. an ultra-thin SiC film on top of AlN. The SiC film, as shown in Fig. 1a of the main text, supports not only the symmetric thin film mode close to $\omega_{LO}$, but additionally an asymmetric thin film mode lying close to the TO frequency $\omega_{TO}$. In the inverse structure, this lower mode disperses at frequencies where the AlN substrate supports a SPhP, and these two modes can also strongly interact. In Fig. \ref{sfig1}b, we show a calculated reflectivity map covering the dispersion range of the substrate AlN SPhP (theoretical curve plotted in dark blue) and the SiC asymmetric thin film mode close to $\omega_{TO}^{SiC}=\wavenumber{796}$ (light blue). In the coupled system, these two modes strongly interact and form an avoided crossing, resulting in two new dispersion branches.

We note that two distinctions have to be made comparing this system with the structure discussed in the main text. First, the dielectric function of SiC close to $\omega_{TO}$ is highly dispersive, and therefore only slightly away from $\omega_{TO}$ the mode loses its ENZ character. Secondly, the material is strongly absorptive at $\omega_{TO}$ and hence the thin-film mode is very likely lossy, hence limiting its practicability. However, such strong coupling in the presence of significant losses is out of the scope of this work.

\section{Deviation from strong coupling for larger film thicknesses}

As stated in the main text, for film thicknesses exceeding $d\gtrsim\nmetr{50}$, the interaction of the modes is not well-described by our strong coupling model anymore. This can be seen both in the mode dispersion predicted by Eq. 2 of the main text and in the field profiles. However, while the latter originates from the approximation of the ENZ mode dispersion, the former provides a physical limit of strong coupling.

The dispersion curves for $d=\nmetr{50,110}$ are shown in Fig. \ref{sfig2}a and b, respectively, plotted on top of numerically calculated reflectivity maps demonstrating a good agreement. However, the increasing deviation of the theoretically predicted dispersion (blue dashed lines) for larger film thicknesses is evident. The reason for that is that in our model, the ENZ mode dispersion is assumed to be constant (\wavenumber{900}), which is sufficiently accurate for $d_{AlN} \lesssim \nmetr{50}$. However, for increasing film thickness the dispersion of the ENZ mode evolves into a thin film mode, see Fig. 1a of the main text. The deviation of the predicted mode dispersion for larger film thicknesses is thus given by the limitation of our model. 

Nonetheless, for increasing film thicknesses, a bare AlN SPhP evolves from the lower mode, which is already largely decoupled from the substrate SiC SPhP at $d\approx\nmetr{110}$. The upper mode, on the other hand, develops towards a pure SiC/AlN interface phonon polariton (IPhP), spectrally confined by the $\omega_{LO}^{AlN}$ frequency and the SiC cut-off frequency $\omega_S^{SiC}$. 

In order to show this disintegration into two distinct modes and thus the breakdown of strong coupling for $d_{AlN} \gtrsim \nmetr{50}$, we analyze the electric field distributions at film thickness $d_{AlN} = \nmetr{110}$. In Fig. \ref{sfig2}c, the normalized in-plane ($x$) electric field amplitude is plotted as a function of frequency and $z$-position (perpendicular to the interfaces) along the three media (air, AlN, and SiC). Clearly, the two modes appear as strong peaks in the field enhancement at their corresponding resonance frequency. But even though the incidence angle was chosen such that the modes are excited at the avoided crossing ($\theta=\dg{28.5}$, $k_{\parallel}/k_0=1.13$), the field distributions of the two modes differ strongly. The lower mode ($\sim\wavenumber{865}$) features the strongest $E_x$ at the air/AlN interface, while the upper mode ($\sim\wavenumber{915}$) exhibits its maximum field enhancement at the SiC/AlN interface, clearly demonstrating that the hybridization of the strong coupling is lost. This can be attributed to the fact that the losses of AlN grow linearly and thus faster than $g_0 \propto \sqrt{d}$ with increasing film thickness $d$. Furthermore, the localization of the modes at the two different interfaces stays the same on both sides of the avoided crossing (not shown), emphasizing their respective bare AlN SPhP and SiC/AlN IPhP nature across the entire dispersion. In fact, both modes feature a similar in-plane field enhancement while the extreme out-of-plane field enhancement is strongly reduced (not shown), meaning that the ENZ characteristics observed in an ultrathin film ($d<\nmetr{50}$) are not present anymore.

\bibliography{supplMat}